\documentclass[aps,twocolumn,showpacs,amssymb]{revtex4}
\usepackage[all]{xy}
\begin{document}

\title{Naturally light right-handed neutrinos in a 3-3-1 Model }

\author{Alex G. Dias,$^{a}$ C. A. de S. Pires,$^b$ P. S. Rodrigues da Silva$^{b}$}
\affiliation{$^a${\small Instituto de  F\'{\i}sica, Universidade
de S\~ao Paulo,\\ Caixa Postal 66.318, 05315-970,S\~ao Paulo-SP, Brazil}\\
$^b${\small  Departamento de F\'{\i}sica, Universidade Federal da
Para\'{\i}ba, Caixa Postal 5008, 58051-970, Jo\~ao Pessoa - PB,
Brazil.}}

\date{\today}
\begin{abstract}
In this work we show  that light right-handed neutrinos, with mass in the sub-eV scale, is a natural outcome in a 3-3-1 model. By considering effective dimension five operators, the model predicts  three light right-handed neutrinos, weakly mixed with the left-handed ones. We show also that the model is able to explain the LSND experiment and still be in agreement with solar and atmospheric data for neutrino oscillation. 
\end{abstract}
\pacs{  }
\maketitle

\section{Introduction}
\label{sec1}
With the exception of  nonzero neutrino mass and mixing,  all the other collected experimental data in particle physics are consistent with the predictions of the standard model of the electroweak and strong interactions~\cite{eidelman}. Concerning neutrinos, the understanding of the smallness of their masses and the largeness of their  mixing, dictated by the neutrino oscillations experiments~\cite{atmos,solar}, is a real puzzle in particle physics at the present. 

On the theoretical side, if only left-handed neutrinos exist, then   the most economical way they can acquire small masses is through  the effective dimension-five  operator~\cite{weinberg}
\begin{eqnarray}
\mathcal{L}_{MP} &=&	\frac{f_{ab}}{\Lambda}(\Phi_l \overline{L^C}_{a m})
(\Phi_n L_{b p})\epsilon_{lm}\epsilon_{np}+\mbox{H.c},
\label{standard5Doperator}
\end{eqnarray}
where $\Phi = \left( \phi^+\,\,,\,\,\phi^0 \right)^T$ and $ L_a=\left( \nu_{aL} \,\,,\,\, e_{aL}\right)^T$. Thus when $\phi^0$ develops a nonzero vacuum expectation value~(VEV), $v$, left-handed neutrinos  automatically develop  Majorana mass 
\begin{eqnarray}
(M_L)_{ab}\overline{(\nu_{aL})^C}\nu_{bL},\,\,\,\,\mbox{with}\,\,\,\,(M_L)_{ab}=\frac{f_{ab}v^2}{\Lambda}.
\label{lightleft}
\end{eqnarray}
 On supposing that such effective operator is realized in some high energy  GUT scale, then we naturally have light masses for the left-handed neutrinos.    

It could be that right-handed neutrinos also exist and they could be light too, but this feature is not naturally obtained in  the standard  model. For instance, to generate light right-handed neutrinos in the standard model, by effective dimension-five operators or by any other mechanism, an intricate combination of symmetries is required, usually  accompanied by a considerable increasing in the particle content~\cite{stepos}. 

On the other hand, in the standard model, light right-handed neutrinos  are interesting only if they can explain the LSND experiment~\cite{LSND}. This requires that the light right-handed neutrinos get weakly mixed with the left-handed ones. People usually refer to these light and weakly mixed  right-handed neutrinos as  sterile neutrinos.

In this letter we examine the problem of generating  light right-handed neutrino masses in  the framework of the 3-3-1 model where those neutrinos are already part of the spectrum. Interesting enough, we show that  light right-handed neutrino masses are a natural outcome in this model. Our approach to the subject is through effective dimension-five operators. Basically, we construct all the effective operators allowed by the symmetries and  particle content of the model and show that they yield light Majorana and Dirac mass terms   for the  neutrinos. Consequently, we will have three light active neutrinos and three light sterile ones\cite{coment}, which makes this 3-3-1 model capable of easily  explaining the LSND experiment.
\section{The model}

The model we consider is the 3-3-1 model with right-handed neutrinos~\cite{footpp,phenomenology}. It is one of the possible models allowed by the $SU(3)_C \otimes SU(3)_L \otimes U(1)$  gauge symmetry where  the fermions are distributed in the following representation content. Leptons come in triplets and singlets
\begin{eqnarray}
L_{a} = \left (
\begin{array}{c}
\nu_{aL} \\
e_{aL} \\
(\nu_{aR})^{c}
\end{array}
\right )\sim(1\,,\,3\,,\,-1/3)\,,\,\,\,e_{aR}\,\sim(1,1,-1),
 \end{eqnarray}
where $a = 1,\,2,\, 3$ refers to the three generations. After the spontaneous breaking of the 3-3-1 symmetry to the standard symmetry, the triplet above splits into the standard lepton doublet $L_a=\left( \nu_{aL} \,\,,\,\, e_{aL}\right)^T$ plus the singlet $( \nu_{aR})^C$. Thus this model recovers the standard model with right-handed neutrinos.

It is not a trivial task to generate light masses to the right-handed neutrinos in any simple extension of the standard model. However, in the 3-3-1 model in question, right-handed neutrinos can naturally obtain small masses through effective dimension-five operators. This is due, in part, to the fact that, in the model, the right-handed neutrinos compose, with the left-handed neutrinos,  the same triplet $L$. As we will see in the next section, it is this remarkable feature that turns feasible  the raise of light right-handed neutrinos.

In the quark sector, one generation comes in the triplet and the other two compose
an anti-triplet with the following content,
\begin{eqnarray}
&&Q_{iL} = \left (
\begin{array}{c}
d_{i} \\
-u_{i} \\
d^{\prime}_{i}
\end{array}
\right )_L\sim(3\,,\,\bar{3}\,,\,0)\,\,\,,u_{iR}\,\sim(3,1,2/3),\,\,\,\nonumber \\
&&\,\,d_{iR}\,\sim(3,1,-1/3)\,,\,\,\,\, d^{\prime}_{iR}\,\sim(3,1,-1/3),\nonumber \\
&&Q_{3L} = \left (
\begin{array}{c}
u_{3} \\
d_{3} \\
u^{\prime}_{3}
\end{array}
\right )_L\sim(3\,,\,3\,,\,1/3)\,,\,\,\,u_{3R}\,\sim(3,1,2/3),\nonumber \\
&&\,\,d_{3R}\,\sim(3,1,-1/3)\,,\,u^{\prime}_{3R}\,\sim(3,1,2/3)
\label{quarks} 
\end{eqnarray}
where $i=1,2$. The primed quarks
are the exotic ones but with the usual electric charges.

In the gauge sector, the model recovers the standard gauge bosons  and disposes of five other gauge bosons called  $V^{\pm}$, $U^0$, $U^{0 \dagger}$ and $Z^{\prime}$~\cite{footpp,phenomenology}. 
Also, the model possesses three scalar triplets, two of them transforming as,  $\eta \sim ({\bf 1}\,,\,{\bf 3}\,,\,-1/3)$ and $\chi \sim ({\bf 1}\,,\,{\bf 3}\,,\,-1/3)$
and the other as, $\rho \sim ({\bf 1}\,,\,{\bf 3}\,,\,2/3)$, with the following vacuum structure~\cite{footpp}
\begin{eqnarray}
\langle \eta \rangle_0 = \left (
\begin{array}{c}
 \frac{v_\eta}{\sqrt{2}} \\
 0\\
 0 
\end{array}
\right ),\,\langle \rho \rangle_0 =\left (
\begin{array}{c}
0 \\
\frac{v_\rho}{\sqrt{2}} \\
0
\end{array}
\right ) ,\, \langle \chi \rangle_0 = \left (
\begin{array}{c}
0 \\
0 \\
\frac{v_{\chi^{\prime}}}{\sqrt{2}}
\end{array}
\right )\,. \label{VEVstructure} 
\end{eqnarray}
These scalars are sufficient to engender spontaneous symmetry breaking and generate the correct masses for all massive particles.

In order to have the minimal model, we assume the following discrete symmetry transformation over the full Lagrangian
\begin{eqnarray}
&&\left( \chi,\,\eta,\,\rho,e_{aR},\,u_{aR},\,u^{\prime}_{3R},\,d_{aR},\,d^{\prime}_{iR}\right) \rightarrow\nonumber \\
 &&-\left( \chi,\,\eta,\,\rho,e_{aR},\, u_{aR},\,u^{\prime}_{3R},\,d_{aR},\,d^{\prime}_{iR}\right),
\nonumber \\
\label{discretesymmetryI}
\end{eqnarray}
where $a=1,2,3$  and $i=1,2$. This discrete symmetry helps in avoiding  unwanted Dirac mass term for the neutrinos~\cite{footpp} and implies a realistic minimal  potential~\cite{pal}. 

With this at hand, the model ends up with the following Yukawa interactions,
\begin{eqnarray}
{\mathcal L}^Y &=&\lambda^1_{ij}\bar Q_{iL}\chi^* d^{\prime}_{jR}+\lambda^2_{33}\bar Q_{3L}\chi u^{\prime}_{3R}+ \lambda^3_{ia}\bar Q_{iL}\eta^* d_{aR}+\nonumber \\
&&\lambda^4_{3a}\bar Q_{3L}\eta u_{aR}+ \lambda^5_{ia}\bar Q_{iL}\rho^* u_{aR}+\lambda^6_{3a}\bar Q_{3L}\rho d_{aR}+\nonumber \\
&& G_{ab}\bar f_{aL} \rho e_{bR}+\mbox{H.c},
\label{yukawa}
\end{eqnarray}
which generate masses for all fermions, with the exception of neutrinos.  
\section{ Neutrino Masses}
 In this section we construct all possible effective dimension-five operators in the 3-3-1 model with right-handed neutrinos that lead to  neutrino masses. The first one involves the triplets $L$ and $\eta$. With these triplets we can form the following  effective dimension-five operator
\begin{eqnarray}
{\mathcal L}_{M_L}&&=	\frac{f_{ab}}{\Lambda}\left( \overline{L^C_a} \eta^* \right)\left( \eta^{\dagger}L_{b} \right)+\mbox{H.c}.
	\label{5dL}
\end{eqnarray}	
According to this operator, when $\eta^0$ develops a VEV, $v_\eta$, the left-handed neutrinos develop  Majorana mass terms with the same form as in Eq. (\ref{lightleft}) but now with
\begin{eqnarray}
(M_{L})_{ab}=\frac{f_{ab}v^2_\eta}{\Lambda}.
	\label{nuL}
\end{eqnarray}

Due to the fact that right-handed neutrinos are not singlets in the model in question, a second  effective dimension-five operator generating neutrino masses is possible. It is constructed with the scalar triplet $\chi$ and the lepton triplet $L$,
\begin{eqnarray}
{\mathcal L}_{M_R}&&=	\frac{h_{ab}}{\Lambda}\left( \overline{L_{a}^C} \chi^* \right)\left( \chi^{\dagger}L_{b} \right)+\mbox{H.c}.
	\label{5dR}
\end{eqnarray}	
When $\chi^{\prime 0}$ develops a VEV, $v_{\chi^{\prime}}$, this effective operator  provides  Majorana masses for the right-handed neutrinos,
\begin{eqnarray}
(M_{R})_{ab}\overline{(\nu_{aR})^C} \nu_{bR},\,\,\,\,\mbox{with} \,\,\,\,(M_{R})_{ab}=\frac{h_{ab}v^2_{\chi^{\prime}}}{\Lambda}
	\label{nuR}
\end{eqnarray}  

Remarkably, a third effective dimension-five operator generating neutrino mass is possible, but now  involving the scalar triplets $\eta$ and $\chi$,
 \begin{eqnarray}
{\mathcal L}_{M_D}&&=	\frac{g_{ab}}{\Lambda}\left( \overline{L_{a}^C}\chi^*\right)\left( \eta^{\dagger}L_{b} \right)+\mbox{H.c},
	\label{5dD}
\end{eqnarray}	
which, remarkably, leads to the following Dirac mass term for the neutrinos,
\begin{eqnarray}
(M_{D})_{ab}\bar \nu_{aR} \nu_{bL},\,\,\,\, \mbox{with} \,\,\,\, (M_{D})_{ab}=\frac{g_{ab}v_{\chi^{\prime}}v_\eta}{\Lambda}.
	\label{nuD}
\end{eqnarray}

Thus, we have Majorana and Dirac mass terms for the neutrinos  both having the same origin, i.e.,  effective dimension-five operators. As in the standard case, on supposing that the three effective dimension-five operators above are realized in some   high energy GUT scale,  we  have thus light Dirac and Majorana mass terms. 

  In view of these neutrino mass terms, the usual manner of proceeding here is to arrange  $M_L$ , $M_R$  and $M_D$  in the following   $6\times 6$ matrix, 
\begin{eqnarray}
\left( \bar{\nu_L^C }\,\,,\,\, \bar \nu_R \right) {\mathcal M} \left (
\begin{array}{c}
\nu_L \\
\nu_R^C 
\end{array}
\right ),
 \label{matrix}
 \end{eqnarray}
in the basis 
\begin{eqnarray}
\left( \nu_L\,,\, \nu^C_R \right)=\left( \nu_{eL}\,,\,\nu_{\mu L}\,,\,\nu_{\tau L}\,,\,\nu^C_{eR}\,,\,\nu^C_{\mu R}\,,\,\nu^C_{\tau R} \right),
\end{eqnarray}
 with
\begin{eqnarray}
{\mathcal M}=\left(\begin{array}{cc}
 M_L & M_D \\
 M^T_D & M_R 
\end{array}
\right).
\label{seessaw}
\end{eqnarray}

At this point, two comments are in order. First, as the VEV $v_{\chi^{\prime}}$ is responsible for the breaking of the 3-3-1 symmetry to the standard symmetry, and that  $v_\eta$ contributes to the spontaneous breaking of the standard symmetry, thus it is natural to expect that $v_{\chi^{\prime}}> v_\eta$, which implies $M_R>M_D>M_L$.  This hierarchy among $M_R$, $M_D$  and $M_L$ leads to a feeble  mixing among the left and right-handed neutrinos, characterizing the last as sterile neutrinos required to explain LSND experiment. Second, the model leads inevitably to three sterile neutrinos.

In order to check this, let us consider the case of one generation. In the basis  $\left( \nu_{eL}\,\,\,\nu_{eR}^C \right)$  we have the mas matrix
\begin{eqnarray} \frac{1}{\Lambda}\left(\begin{array}{cc}
 fv^2_\eta & g v_\eta v_{\chi^{\prime}} \\
 gv_\eta v_{\chi^{\prime}} &  h v^2_{\chi^{\prime}}
\end{array}
\right).
 \label{onegenera}
 \end{eqnarray}
 
By diagonalizing this matrix  for $v_{\chi^{\prime}}> v_\eta$, we obtain the eigenvalues
\begin{eqnarray}
	\frac{fh-g^2}{h}\frac{v^2_\eta}{\Lambda} \,\,\,,\,\,\, h\frac{v^2_{\chi^{\prime}}}{\Lambda},
	\label{eigenvalues}
\end{eqnarray}
and the correspondent eigenvectors
\begin{eqnarray}
	&&N_1=\nu_{eL}+\frac{fh-g^2}{gh}\frac{v_\eta }{v_{\chi^{\prime}}}(\nu_R)^C,\nonumber \\
	&& N_2=(\nu_R)^C-\frac{fh-g^2}{gh}\frac{v_\eta }{v_{\chi^{\prime}}}\nu_{eL}.
	\label{eigenvectors}
\end{eqnarray}

We see that the magnitude of the mixing is basically established by the VEV´s $v_\eta$ and $v_{\chi^{\prime}}$  through the ratio $\frac{v_\eta }{v_{\chi^{\prime}}}$. The typical values of such VEV´s are $v_\eta =10^2$~GeV and $v_{\chi^{\prime}}=10^3$~GeV. This leads to an active-sterile mixing of order of $10^{-1}$ which falls in the expected range of values required to explain LSND as discussed below. 

In order to explain LSND experiment, we need at least one sterile neutrino. In the 3-3-1 model with right-handed neutrinos we have necessarily three sterile neutrinos. The masses and mixing of the neutrinos is dictated by the matrix ${\mathcal M}$ in Eq. (\ref{seessaw}). As in the case of quarks and charged leptons, the masses and mixing angles of the active and sterile neutrinos is a question of an appropriate tunning of the couplings $f_{ab}$ , $g_{ab}$  and $h_{ab}$. 

Presently we have three kinds of experimental evidence for neutrino oscillation. One  involves  neutrino oscillation from the sun whose data are~\cite{atmos},
\begin{eqnarray}
90\% \mbox{CL}\,\,\,	1.5 \times 10^{-3} \mbox{eV}^2 &\leq& \Delta m^2_{23}\leq 3.4 \times 10^{-3}\mbox{eV}^2\,,
\nonumber \\
\sin^2 2 \theta_{23} &&> 0.92.
\label{atmosevid}
\end{eqnarray}
while the other  evidence involves solar neutrino oscillation. The data in this case are~\cite{solar},
\begin{eqnarray}
90\% \mbox{CL}\,\,	7.4 \times 10^{-5} \mbox{eV}^2 &\leq& \Delta m^2_{12}\leq 8.5 \times 8.5\times 10^{-5}\mbox{eV}^2\,, \nonumber \\
0.33 &\leq& \tan^2 \theta_{12} \leq 0.50,
\label{solarevid}
\end{eqnarray}

The third evidence refers to the appearance of $\bar \nu_e$  in a beam of $\bar \nu_\mu$ observed by the LSND experiment~\cite{LSND}. This experiment does not have the status of the solar and atmospheric ones, since it needs to be confirmed. The Mini Boone experiment is in charge of this~\cite{miniboone}. 
 
The analysis of the data from LSND depends on the number of sterile neutrinos we suppose. For the case of only one sterile neutrino we have, the so called $3+1$ scenario, where~\cite{oneste},
\begin{eqnarray}
	\Delta m^2_{41} = 0.92 \mbox{eV}^2\,,\,\,\, U_{e 4}=0.136 \,\,\, \mbox{and}\,\,\,  U_{\mu 4}=0.205\,.
	\label{LSND31} 
\end{eqnarray}

According to Ref.~\cite{sorel}, for the case of two sterile neutrinos, called $3+2$ scenario~\cite{twoste}, we can have two possible schemes. In one case, the best fit leads to
\begin{eqnarray}
&&	\Delta m^2_{41} = 0.92 \mbox{eV}^2\,,\,\,\, U_{e 4}=0.121 \,\,\, \mbox{and}\,\,\,  U_{\mu 4}=0.204,\nonumber \\
&& \Delta m^2_{51} = 22 \mbox{eV}^2\,,\,\,\, U_{e 5}=0.036 \,\,\, \mbox{and}\,\,\,  U_{\mu 5}=0.224\,,
	\label{LSND321} 
\end{eqnarray}
in the other case, we have
\begin{eqnarray}
&&	\Delta m^2_{41} = 0.46 \mbox{eV}^2\,,\,\,\, U_{e 4}=0.090 \,\,\, \mbox{and}\,\,\,  U_{\mu 4}=0.226,\nonumber \\
&& \Delta m^2_{51} = 0.89 \mbox{eV}^2\,,\,\,\, U_{e 5}=0.125 \,\,\, \mbox{and}\,\,\,  U_{\mu 5}=0.160.
	\label{LSND322} 
\end{eqnarray}

We would like to provide a texture for ${\mathcal M}$ that solves neutrino oscillation, i.e., that recovers as close as possible the neutrino data showed above. But as the 3-3-1 model provides  three sterile neutrinos and we dispose of  analysis considering at most two sterile neutrinos, we have to make some assumptions. We will neglect CP violation; assume that the third sterile neutrino decouples from the others; take $U_{e3}=0$ and consider that the atmospheric angle is exactly maximal. By an appropriate choice of the free parameters $f_{ab}$, $g_{ab}$  and $h_{ab}$, and taking $v_\eta =10^2$~GeV, $v_{\chi^{\prime}}=10^3$~GeV  and $\Lambda=10^{14}$~GeV, a possible  texture for  ${\mathcal M}$ that incorporates  such assumptions is,
\begin{eqnarray}\begin{small}
{\mathcal M}=	\left[ \begin {array}{cccccc}  0.0465& 0.0208&- 0.0208& 0.121& 0.136& 0.0\\\noalign{\medskip}
 0.0208& 0.064&- 0.0166&- 0.0495&
 0.167& 0.0\\\noalign{\medskip}- 0.0208&- 0.0166&
 0.064& 0.0495&- 0.167& 0.0\\\noalign{\medskip}
 0.121&- 0.0495& 0.0495& 0.66& 0.0& 0.0
\\\noalign{\medskip} 0.136& 0.167&- 0.167& 0.0&
 0.851& 0.0\\\noalign{\medskip} 0.0& 0.0& 0.0& 0.0& 0.0&
 1.7\end {array} \right]\end{small}
(\mbox{eV}) \,.
\label{texture}
\nonumber \\
\end{eqnarray}
This mass matrix is diagonalized by the following mixing matrix,
\begin{eqnarray}\begin{small}
U^{(6)}\approx 
\left[ \begin {array}{cccccc}  0.847& 0.476& 0.0& 0.179& 0.154& 0.0\\\noalign{\medskip}- 0.344& 0.581& 0.71&- 0.0733& 0.189& 0.0
\\\noalign{\medskip} 0.344&- 0.581& 0.71& 0.0733&- 0.189& 0.0
\\\noalign{\medskip}- 0.207& 0.0& 0.0& 0.978& 0.0& 0.0
\\\noalign{\medskip} 0.0&- 0.309& 0.0& 0.0& 0.951& 0.0
\\\noalign{\medskip} 0.0& 0.0& 0.0& 0.0& 0.0& 1.0\end {array} \right]\end{small}
\label{mixing}
\end{eqnarray}

which leads to the following neutrino masses,
\begin{eqnarray}
&& m_1\approx-5.5\times 10^{-5}\mbox{eV},\,\, m_2\approx 9.3\times 10^{-3}\mbox{eV},\nonumber \\
&& m_3\approx4.8\times 10^{-2}\mbox{eV}\,\,, m_4\approx6.9\times 10^{-1}\mbox{eV},\nonumber \\
&&m_5\approx9.4\times 10^{-1}\mbox{eV},\,\,m_6=1.7\mbox{eV}\,.
	\label{numasses6}
\end{eqnarray}
The values for the neutrino masses in Eq.~(\ref{numasses6}) and the pattern of $U^{(6)}$ above, Eq.~(\ref{mixing}), accommodate the solar and atmospheric oscillation data along with the  LSND experiment altogether. For the sterile neutrinos, we did not recover exactly the best fit, which we believe is due only to the set of assumptions made above.   

We also would like to call the  attention to the fact that our  sterile neutrinos present  non-standard interactions. For example, they interact directly to the charged leptons through a new charged gauge boson according to
\begin{eqnarray}
	\frac{g}{\sqrt{2}}(\bar{e}_L)^C \gamma^\mu \nu_R V^+_\mu +\mbox{H.c}\,,
	\label{leptonsterile}
\end{eqnarray}
and also couple to the active neutrinos through a new non-hermitian neutral gauge boson according to,
\begin{eqnarray}
\frac{g}{\sqrt{2}}(\bar{\nu}_L)^C \gamma^\mu \nu_R U^{0 \dagger}_\mu +\mbox{H.c}\,.
\label{activesterileint}
\end{eqnarray}
This turns the phenomenology of our sterile neutrinos much richer than usual. For example, in the standard case there is a conflict among cosmology and LSND result~\cite{lsndcosm}. We are not sure if such conflict will persist here in view of these non-standard interactions. This means that our sterile neutrinos scenario demands that their  cosmological aspects as, for instance, their abundance and temperature decoupling~\cite{lsndcosm}, their contribution to the dark matter~\cite{darkmatter} and, mainly, their implications for the big bang nucleosynthesis~\cite{BBN}, must be revisited in the light of the non-standard interactions just mentioned\cite{newways}. 

Finally, we would like to say that although our sterile neutrinos couple directly to the active ones, see Eq.~(\ref{activesterileint}), they are still stable. The interactions in Eq.~(\ref{activesterileint}) allows the  decay of the heavier sterile neutrinos in lighter neutrinos. For example we can have the following channel $\nu_{\tau R} \rightarrow \nu_{\tau L}\nu_{e L}\nu_{eR}$. In this case we have the following expression for the decay width
\begin{eqnarray}
	\Gamma=G^2_F\frac{m^5_{\nu_{\tau R}}m^4_W}{192\pi^3 m^4_U}.
	\label{decay}
\end{eqnarray}
For $m_U=250$~GeV and $m_{\nu_{\tau R}}=1.5$~eV, we obtain a life-time of order of $3.4\times10^{34}$s, leading, thus, to  stable sterile neutrinos.  

\section{Conclusions}
The main achievement of this work is to show that the right-handed neutrinos that appear in a version of the 3-3-1 model can be naturally light when dimension five effective operators are included. Such neutrinos can be identified as sterile ones, offering us the possibility of explaining the LSND data. We have checked that, 
although the model leads to a $3+3$ scenario, the results from a $3+2$ scenario can be easily recovered by making an appropriate choice of the Yukawa couplings in the mass matrix ${\mathcal M}$.
Remarkably, we have shown that besides LSND, this 3-3-1 model has all the necessary features to also explain solar and atmospheric neutrino oscillation data without adding any extra fields or intricate symmetries. 
This is an automatic property as far as the allowed dimension five operators here included can be embedded in some larger underlying theory, maybe a GUT or something else at higher energies than TeV scale.

Moreover, it is possible that the new non-standard interactions involving neutrinos can reveal a very different perspective concerning the conflict between LSND results and neutrino cosmology. That is something to be further analyzed but it is out of the scope of this work. However, we should stress that all this characteristics of this 3-3-1 model are fairly appealing, considering the tiny amount of assumptions we had to rely on.

\vspace{1.0cm} \noindent {\bf Acknowledgments}

\noindent 
We thanks O. G. Peres and V. Pleitez for useful discussion about sterile neutrinos phenomenology. This work was supported by Conselho Nacional de Pesquisa e
Desenvolvimento - CNPq(CASP,PSRS),  Funda\c{c}\~ao de Amparo \`a Pesquisa do Estado 
de S\~ao Paulo - FAPESP(AGD) and by Funda\c c\~ao de
Apoio \`a Pesquisa do Estado da Para\'{\i}ba(FAPESQ)(CASP).



\begin{thebibliography}{99}
\bibitem{coment}
Although we refer to those neutrinos as sterile neutrinos, we call the attention to the fact  that in the framework of the 3-3-1 model those neutrinos are not sterile because they interact with the active neutrinos and the charged leptons as showed in Eqs. (\ref{leptonsterile})  and (\ref{activesterileint}).
\end{thebibliography}
\end{document}